\documentclass[10pt,letterpaper]{article}
\usepackage{opex3}

\begin{document}

\title{Parametric Frequency Conversion of Short Optical Pulses Controlled by a CW Background}

\author{Matteo Conforti, Fabio Baronio}

\address{Dipartimento di Elettronica per l'Automazione,Universit\`a di Brescia, Via Branze 38, 25123 Brescia, Italy}

\email{fabio.baronio@ing.unibs.it} 

\author{Antonio Degasperis}

\address{Dipartimento di Fisica, Istituto Nazionale di Fisica Nucleare, Universit\`a ``La Sapienza'', P.le A. Moro 2, 00185 Roma, Italy}

\author{Stefan Wabnitz}

\address{Institut Carnot de Bourgogne, UMR 5209 CNRS,Universit\'e de Bourgogne, 9 Av. A. Savary BP 46870, 21078 Dijon, France}



\begin{abstract}
We predict that parametric sum-frequency generation of an
ultra-short pulse may result from the mixing of an ultra-short
optical pulse with a quasi-continuous wave control. We
analytically show that the intensity, time duration and group
velocity of the generated idler pulse may be controlled in a
stable manner by adjusting the intensity level of the background
pump.
\end{abstract}



\section{Introduction}
Optical parametric amplification in quadratic nonlinear crystals
has been studied since the invention of the laser, as it provides
a versatile means of achieving widely tunable frequency conversion
\cite{cerullo03}. In parametric processes, the effective
interaction length of short optical pulses is limited by temporal
walk-off owing to chromatic dispersion, or group velocity mismatch
(GVM) \cite{armst70,Amanov92}. Compression and amplification of
ultra-short laser pulses in second harmonic and sum-frequency (SF)
generation in the presence of GVM was theoretically predicted
\cite{wang90,stabinis91} and observed in several experiments
\cite{wang92,chien94}. The conversion efficiency of generated SF
pulses may be optimised
\cite{ibragimov96,ibragimov97,ibragimov98,ibragimov99,fournier98}
by operating in the soliton regime \cite{zakharov73,kaup76}. In
fact, the temporal collision of two short soliton pulses in a
quadratic nonlinear crystal may efficiently generate a short,
time-compressed SF pulse \cite{ibragimov96}. However this SF pulse
is unstable: its energy decays back into the two incident pulses
after a relatively short distance.

In this Paper we consider the parametric SF conversion from the
mixing of an ultra-short signal pulse with a quasi-continuous wave
(CW) or background pump, in the presence of GVM. Quite
surprisingly we find that parametric mixing of these waves may
lead to highly efficient generation of stable and ultra-short
idler pulses. Indeed, we predict that the interaction of an
ultra-short signal with a CW pump may generate a stable three-wave
resonant interaction simulton (TWRIS) \cite{noza73,noza74}, consisting of a locked
bright-bright-dark triplet (signal-idler-pump) that propagates
with a single nonlinear velocity \cite{cal05,deg06}. We analytically
show that the intensity, time duration and group velocity of the
generated idler pulse may be controlled in a stable manner \cite{conforti06} by
means of simply adjusting the intensity level of the CW
background. Although we shall restrict our attention in this work to a travelling-wave
interaction geometry, we may anticipate that our results will have important ramifications in the
optimization of the efficiency of ultrashort pulse optical parametric oscillators \cite{pico98,gale95}.

\section{Three-wave-interaction equations}
The equations describing the quadratic resonant interaction of
three waves in a nonlinear medium read as:

\begin{eqnarray}\label{3wri}
\nonumber \frac{\partial A_{1}}{\partial \xi}+\delta_1
\frac{\partial A_1}{\partial \tau}&=& i A_2^*A_3,\\
 \frac{\partial A_{2}}{\partial \xi}+\delta_2
\frac{\partial A_2}{\partial \tau}&=& i A_1^*A_3,\\ \nonumber
 \frac{\partial A_{3}}{\partial \xi}+\delta_3
\frac{\partial A_3}{\partial \tau}&=& i A_1\, A_2,
\end{eqnarray}
with
\begin{eqnarray}\label{dimensi}
 A_j&=&  \, \pi \chi^{(2)} \,
\sqrt{\frac{n_j\omega_1 \omega_2 \omega_3}{n_1 n_2 n_3 \omega_j}}
\ E_j \, .
\end{eqnarray}
Here $\tau=t/t_0$, $t_0$ is an arbitrary time parameter;
$\xi=z/z_0$, $z_0$ is an unit space-propagation parameter. $E_j$
are the slowly varying electric field envelopes of the waves at
frequencies $\omega_j$, $n_j$ are the refractive indexes,
$\chi^{(2)}$ is the quadratic nonlinear susceptibility, $\delta_j=
z_0/(v_j t_0)$ with $v_j$ the linear group velocities, and
$j=1,2,3$. We assume that the group velocity $v_3$ of the wave
with the highest frequency ($\omega_3=\omega_1+\omega_2$) lies
between the group velocities of the other waves, i.e.
$v_1>v_3>v_2$. With no loss of generality, we shall write the Eqs.
(\ref{3wri}) in a coordinate system such that $\delta_1=0$, which
implies $0<\delta_3<\delta_2$. Eqs. (\ref{3wri}) exhibit the
conserved quantities
\begin{equation}\label{E12}
 U_{13}=U_1+U_3=\frac{1}{2}\int_{-\infty}^{+\infty} ( |A_1|^2 + |A_3|^2) d\tau,
 \end{equation}
\begin{equation}\label{E23}
 U_{23}=U_2+U_3=\frac{1}{2}\int_{-\infty}^{+\infty}( |A_2|^2 + |A_3|^2 )
 d\tau,
 \end{equation}
 \begin{equation}\label{E123}
 U=U_1+U_2+2U_3=\frac{1}{2}\int_{-\infty}^{+\infty}( |A_1|^2+|A_2|^2 + 2|A_3|^2 )
 d\tau.
 \end{equation}
where $U_1$, $U_2$ and $2 U_3$ represent the energies at the
frequencies $\omega_1$, $\omega_2$ and $\omega_3$.
\section{Soliton-based parametric sum-frequency conversion}
\begin{figure}[h]
 \begin{center}
     \includegraphics[width=5cm]{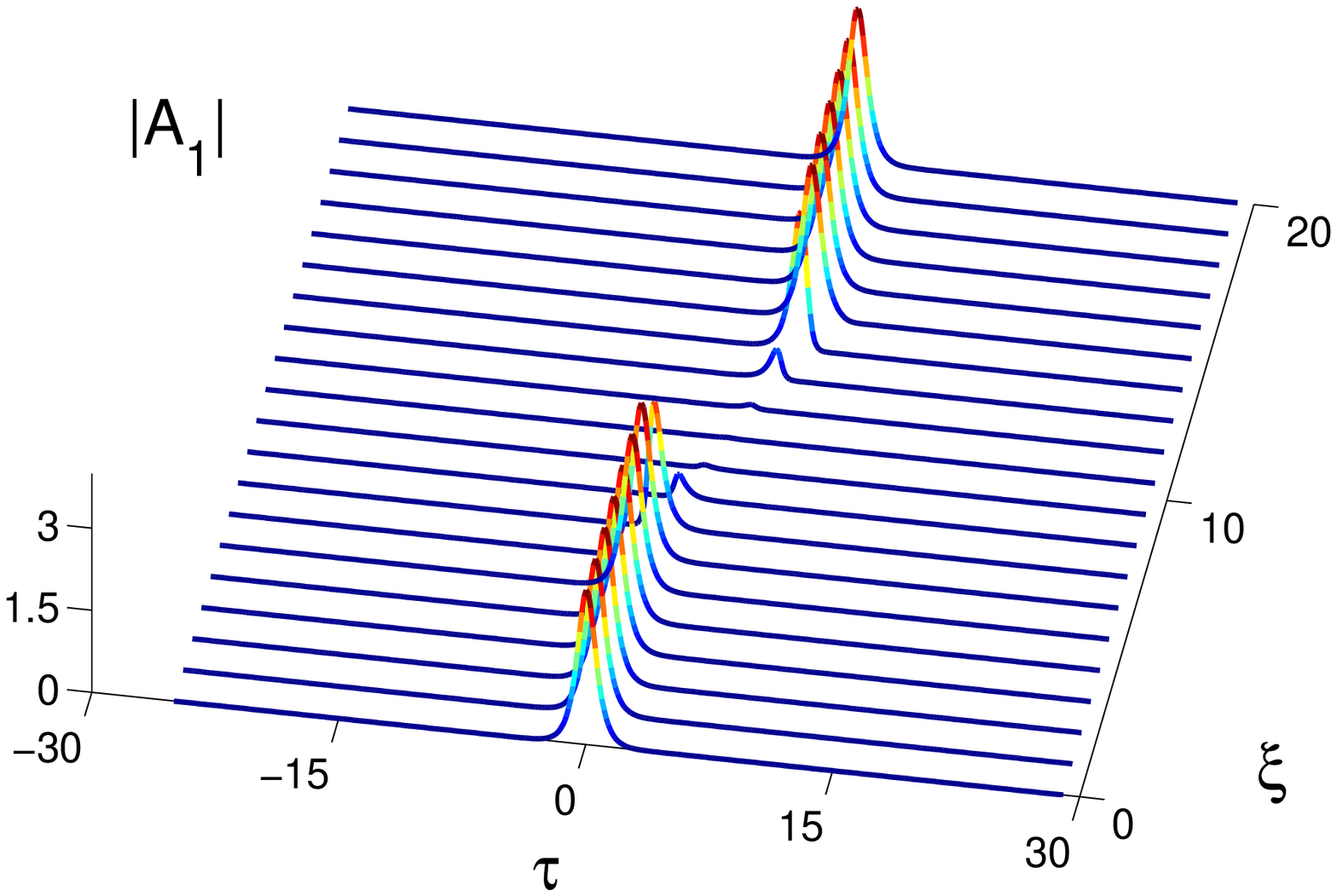}
     \includegraphics[width=5cm]{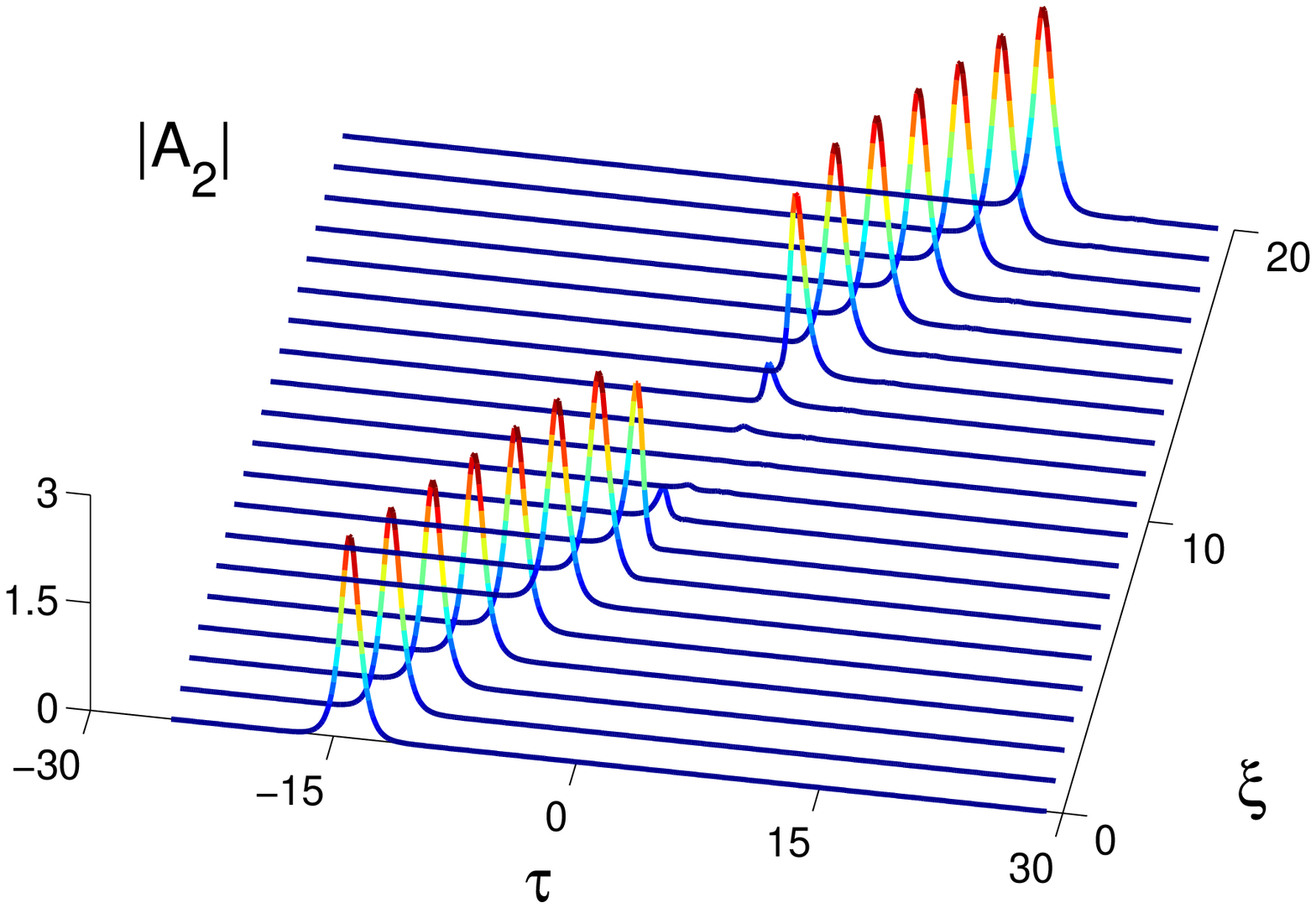}
     \includegraphics[width=5cm]{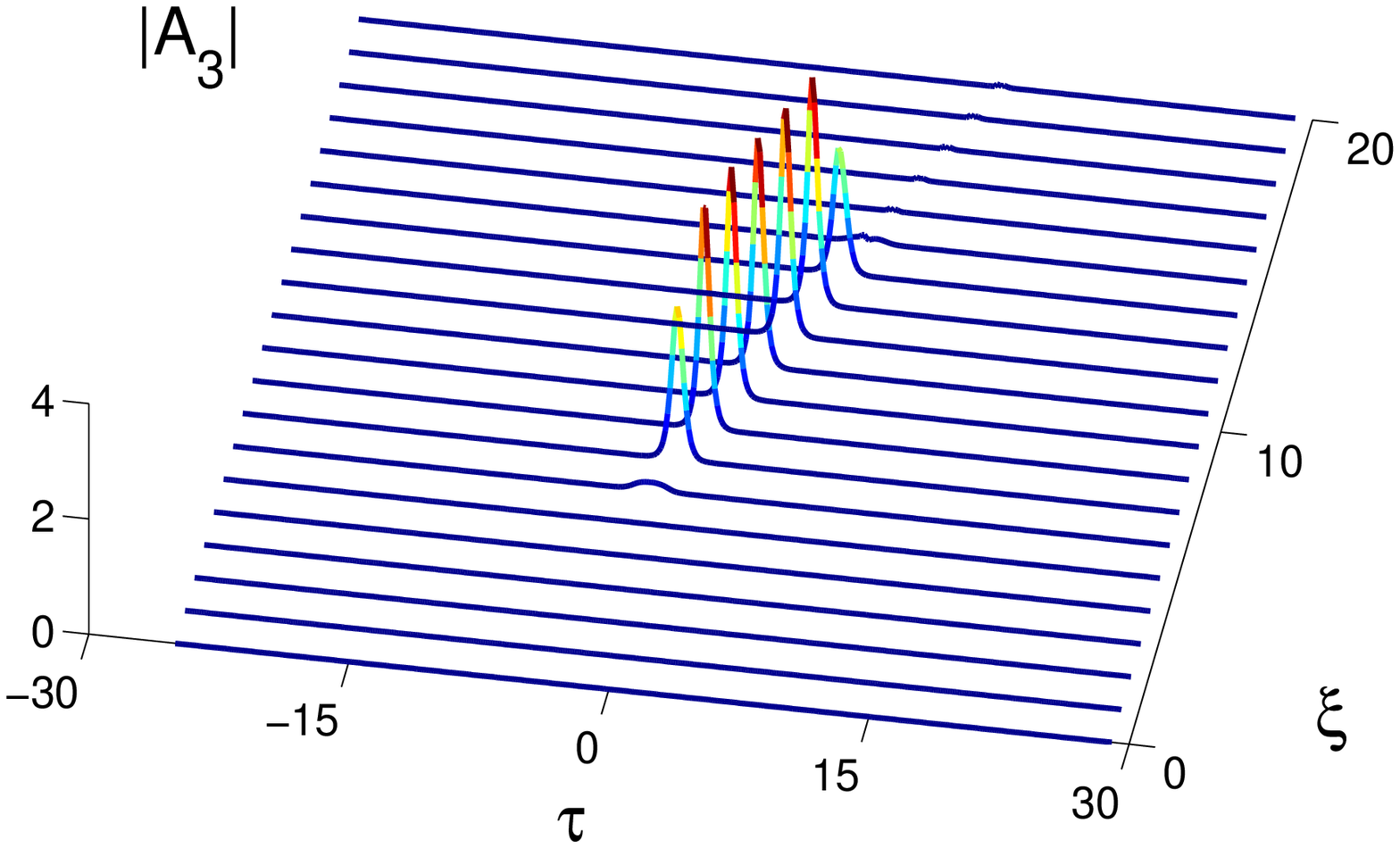}
    \end{center}
     \caption{Sum-frequency parametric interaction of two short
     optical signals at $\omega_1$ and $\omega_2$. The characteristic delays
are $\delta_1=0, \delta_2=2, \delta_3=1$.}\label{bbbnumerics}
\end{figure}

Figure \ref{bbbnumerics} illustrates a typical example of the
efficient SF parametric interaction of two short optical pulses in
the soliton regime \cite{ibragimov96}.

At the crystal input, two isolated pulses $A_1$ and $A_2$ with
frequencies $\omega_1$ and $\omega_2$ propagate with speeds $v_1$
and $v_2$. Whenever the faster pulse overtakes the slower one, an
idler pulse $A_3$ at the SF $\omega_1+\omega_2$ is generated and
propagates with the linear speed $v_3$. Depending on the time
widths and intensities of the input pulses, the duration of the SF
pulse is reduced with respect to the input pulse widths.
Correspondingly, the SF pulse peak intensity grows larger than the
input pulse intensities. Figure \ref{bbbnumerics} shows that,
eventually, the SF idler pulse decays back into the two original
isolated pulses at frequencies $\omega_1$ and $\omega_2$. Note
that the shapes, intensities and widths of the input pulses are
left unchanged in spite of their interaction. As shown in Ref.
\cite{ibragimov96}, the above discussed SF pulse generation
process may be analytically described in terms of soliton
solutions of Eqs. (\ref{3wri}) \cite{zakharov73,kaup76}. The decay
of the SF pulse which is shown in Fig. \ref{bbbnumerics} may be a
significant drawback in practical applications, since it implies
that a given nonlinear crystal length yields efficient conversion
for a limited range of input pulse intensities and time widths
only.
%
\begin{figure}
 \begin{center}
     \includegraphics[width=5cm]{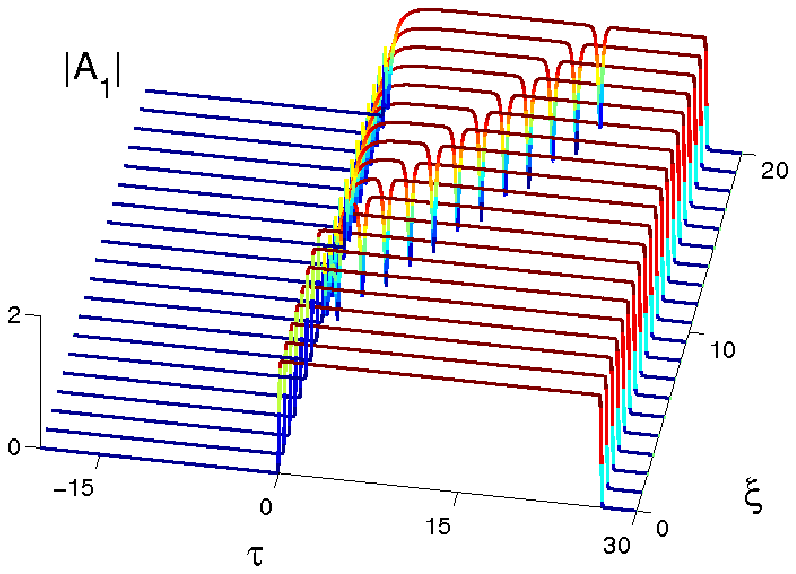}
     \includegraphics[width=5cm]{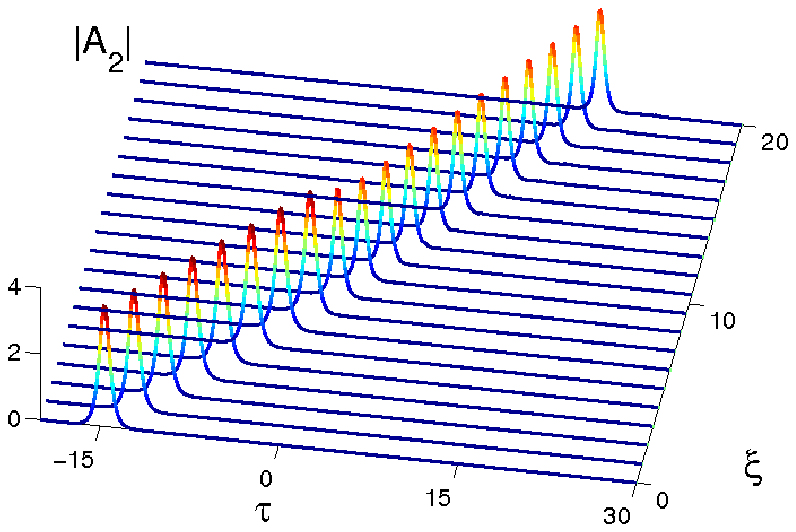}
     \includegraphics[width=5cm]{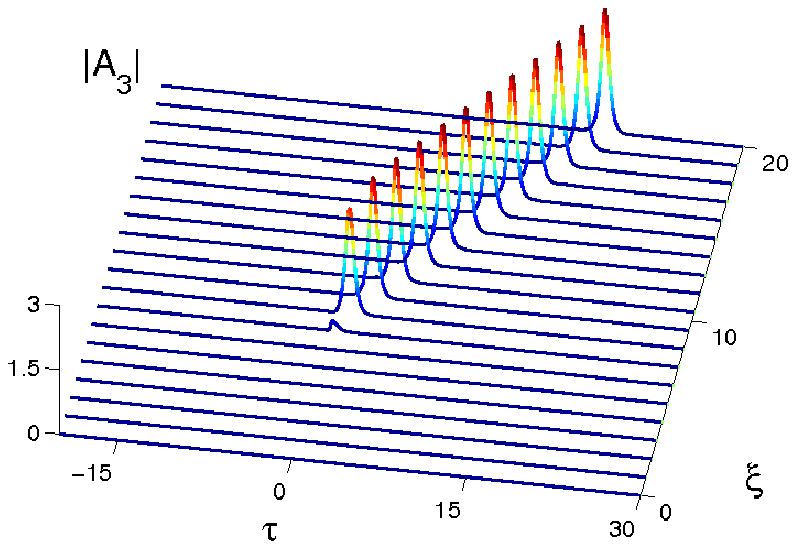}
    \end{center}
     \caption{Sum-frequency parametric interaction of a short
     pulse at $\omega_2$ and a quasi-CW control at $\omega_1$. The characteristic delays
are $\delta_1=0, \delta_2=2, \delta_3=1$.}\label{bbk}
\end{figure}

%

Here we demonstrate that the parametric sum-frequency conversion
of an ultra-short signal and a quasi-CW background pump-control
may be exploited as a means to reduce or even eliminate the decay
of the generated idler wave. In the presence of GVM, the
parametric SF conversion of an ultra-short optical signal and a
quasi-CW pump typically leads to the generation of a low-intensity
and relatively long idler pulse, whose duration is associated with
the interaction distance in the crystal. This scenario changes
dramatically in the soliton regime. Figure \ref{bbk} illustrates
the efficient generation of a stable, ultra-short SF idler pulse
from the parametric SF conversion of a properly prepared
ultra-short signal and an arbitrary intensity level CW background
control.

In Fig. \ref{bbk} we injected in the quadratic nonlinear crystal
the short signal at frequency $\omega_2$, along with a delayed and
relatively long pump-control pulse at frequency $\omega_1$.
Initially, the two pulses propagate uncoupled; as soon as the
faster pulse starts to overlap in time with the slower quasi-CW
control, their nonlinear mixing generates a short SF idler pulse.
The sum-frequency process displayed in Fig. \ref{bbk} can be
analytically explained and explored in terms of stable TWRIS
solutions \cite{deg06}. In the notation of Eqs. (\ref{3wri}), the
TWRIS solution reads as

\begin{eqnarray}\label{simult-}
A_1&=&-\{1+\frac{2p \, b^*}{|b|^2+|a|^2} [1-\tanh[B(\tau+\delta\xi
)]]\} \, \frac{i \, a\, g_3
\exp(iq_3\tau_3)}{g(\delta_2-\delta_3)} \nonumber \\
A_2&=&\frac{-2p \, a^*}{\sqrt{|b|^2+|a|^2}} \frac{i \,
g_1}{g(\delta_2-\delta_3)} \frac{\exp[i(q_1\tau_1-\chi \tau+\omega
\xi )]}{\cosh[B(\tau+\delta\xi )]},  \nonumber \\
A_3&=& \frac{-2p \, b^*}{\sqrt{|b|^2+|a|^2}}\frac{i\,
g_2}{g(\delta_2-\delta_3)} \frac{\exp[-i(q_2\tau_2+ \chi
\tau-\omega \xi)]}{\cosh[B(\tau+\delta\xi )]},
\end{eqnarray}
where
\begin{eqnarray}\label{def}
b &=& (Q -1)(p + i k/Q),\ \ \ \ \ r=p^2-k^2-|a|^2, \nonumber \\
Q &=& \frac{1}{p}\sqrt{ \frac12 [\,\,r + \sqrt{r^2
 +4k^2p^2}\,\,]},\nonumber \\
B &=& p[\,\delta_2+\delta_3 - Q
(\delta_2-\delta_3)\,]/(\delta_2-\delta_3), \nonumber
\\
\delta &=&  2\delta_2 \delta_3/ [\,\delta_2+\delta_3 - Q
(\delta_2-\delta_3)\,],\nonumber
\\
\chi &=& k [\,\delta_2+\delta_3
-(\delta_2-\delta_3)/Q\,]/(\delta_2-\delta_3),\nonumber
\\
\omega &=&  -2k \delta_2 \delta_3/(\delta_2- \delta_3), \tau_n=
\tau+\delta_n \xi \nonumber\\
q_n&=&q(\delta_{n+1}-\delta_{n+2}), \,
g_n=|(\delta_{n}-\delta_{n+1})\,(\delta_{n}-\delta_{n+2})|^{-1/2}\
\nonumber \\
g&=&g_1\,g_2\,g_3\,\,, \, \, \,n=1,2,3 \ mod \, (3).
%
%
\end{eqnarray}
For a given choice of the characteristic linear group velocities,
we are left with the four independent parameters $p,a,k,q$. The
parameter $p$ is associated with the re-scaling of the wave
amplitudes, and of the coordinates $\tau$ and $\xi$. Whereas $a$
measures the amplitude of the CW background in wave $A_1$ (namely
$ a \sqrt{\delta_2\delta_3}$). The value of $k$ is related to the
soliton wave—number. The parameter $q$ simply adds a phase shift
which is linear in both $\tau$ and $\xi$ (see \cite{deg06} for
parameter details).

At the input, the properly prepared short pulse at frequency
$\omega_2$ and with a speed $v_2$ is a stable single component
TWRIS (\ref{simult-}) with parameters $p>0,k,q,a=0$. When this
faster pulse, pre-delayed with respect to the slower quasi-CW pump
at frequency $\omega_1$, overtakes the background (at $\tau=0$, in
Fig. \ref{bbk}), their collision leads to the generation of a
short idler pulse at the SF $\omega_3$. Additionally, a dip
appears in the quasi CW-control; whereas the intensity, duration
and propagation speed of the input wave at frequency $\omega_2$
are modified. Indeed, the signal-pump interaction generates a new
stable TWRIS (\ref{simult-}), with parameters
$\overline{p},\overline{k},\overline{q},\overline{a}$, moving with
the locked nonlinear velocity $\overline{v}=z_0/(t_0\overline{
\delta})$, where $\overline{\delta}$ is given in (\ref{def}).

It is remarkable that we may analytically predict the parameters
$\overline{p},\overline{k},\overline{q},\overline{a}$ of the
generated TWRIS from the corresponding parameters of the input
single wave TWRIS and the complex amplitude of the pump control.
This can be achieved by supposing that the input TWRIS
adiabatically (i.e., without emission of radiation) reshapes into
a new TWRI simulton after its collision with the quasi-CW pump at
a given point in time (say, at $\tau=0$).  Under this basic
hypothesis, the conservative nature of the three-wave interaction
permits us to suppose that: i) the energy $U_{23}$ (\ref{E23}) of
the input TWRI soliton is conserved in the generated TWRI
simulton; ii) the phase of the $\omega_2$ frequency components of
the input TWRI soliton and of the generated TWRI simulton is
continuous across their time interface (i.e., at $\tau=0$); iii)
the amplitude and phase of the control pump $C$ coincide with the
corresponding values of the asymptotic plateau of the generated
TWRI simulton component at frequency $\omega_1$. By imposing the
above three conditions, after some straightforward calculations we
obtain the following relations that relate the parameters of the
incident and of the transmitted TWRIS

\begin{eqnarray}\label{defteor}
\overline{p} &=& p \nonumber \\
\overline{a} &=& |C|/ \sqrt{\delta_2\delta_3} \nonumber \\
\overline{q} &=& \angle(C) \nonumber \\
\overline{k} &=& k+q/2-\overline{q}/2.
\end{eqnarray}

\begin{figure}[h]
 \begin{center}
     \includegraphics[width=3.9cm]{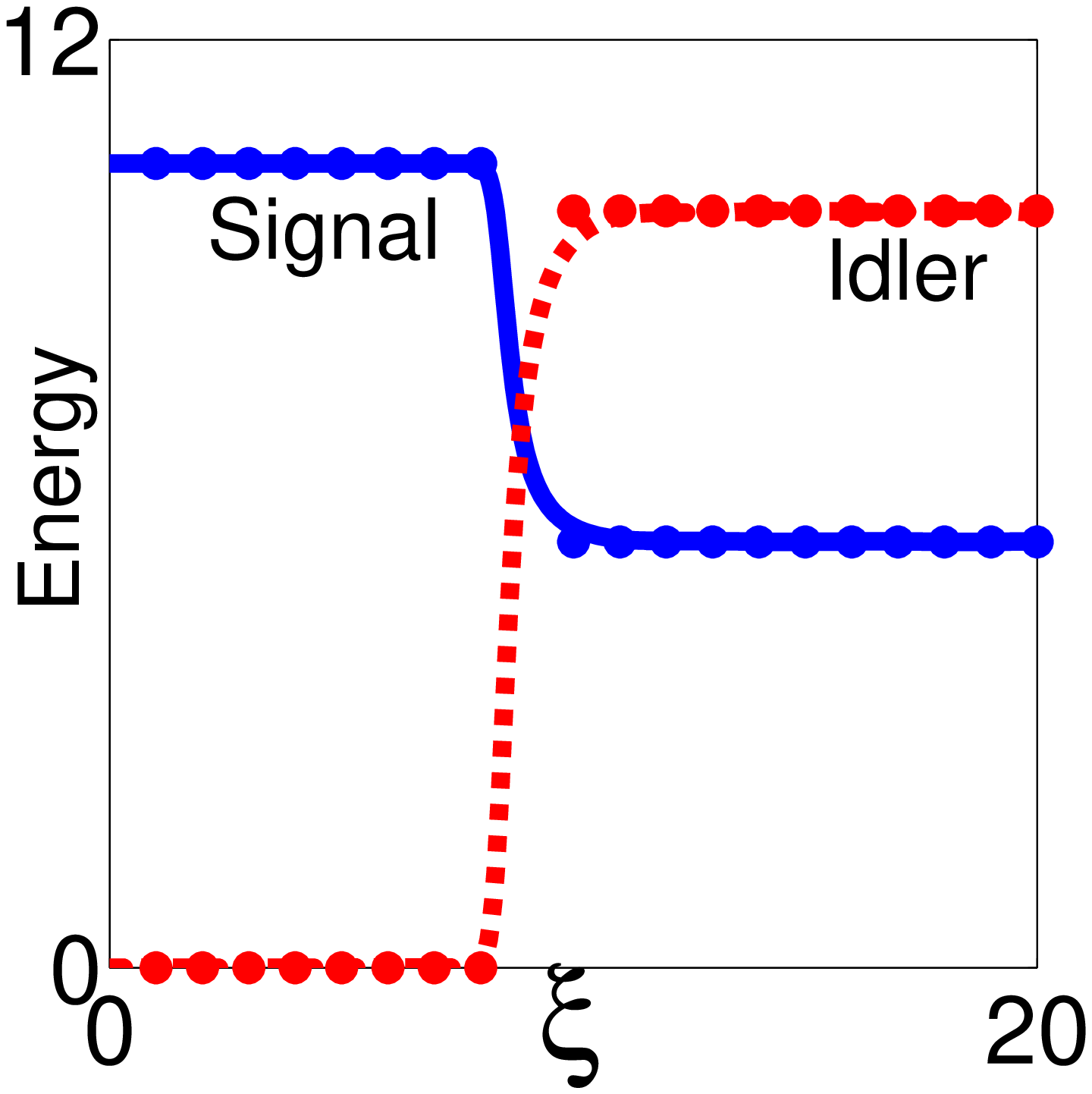}
     \includegraphics[width=3.9cm]{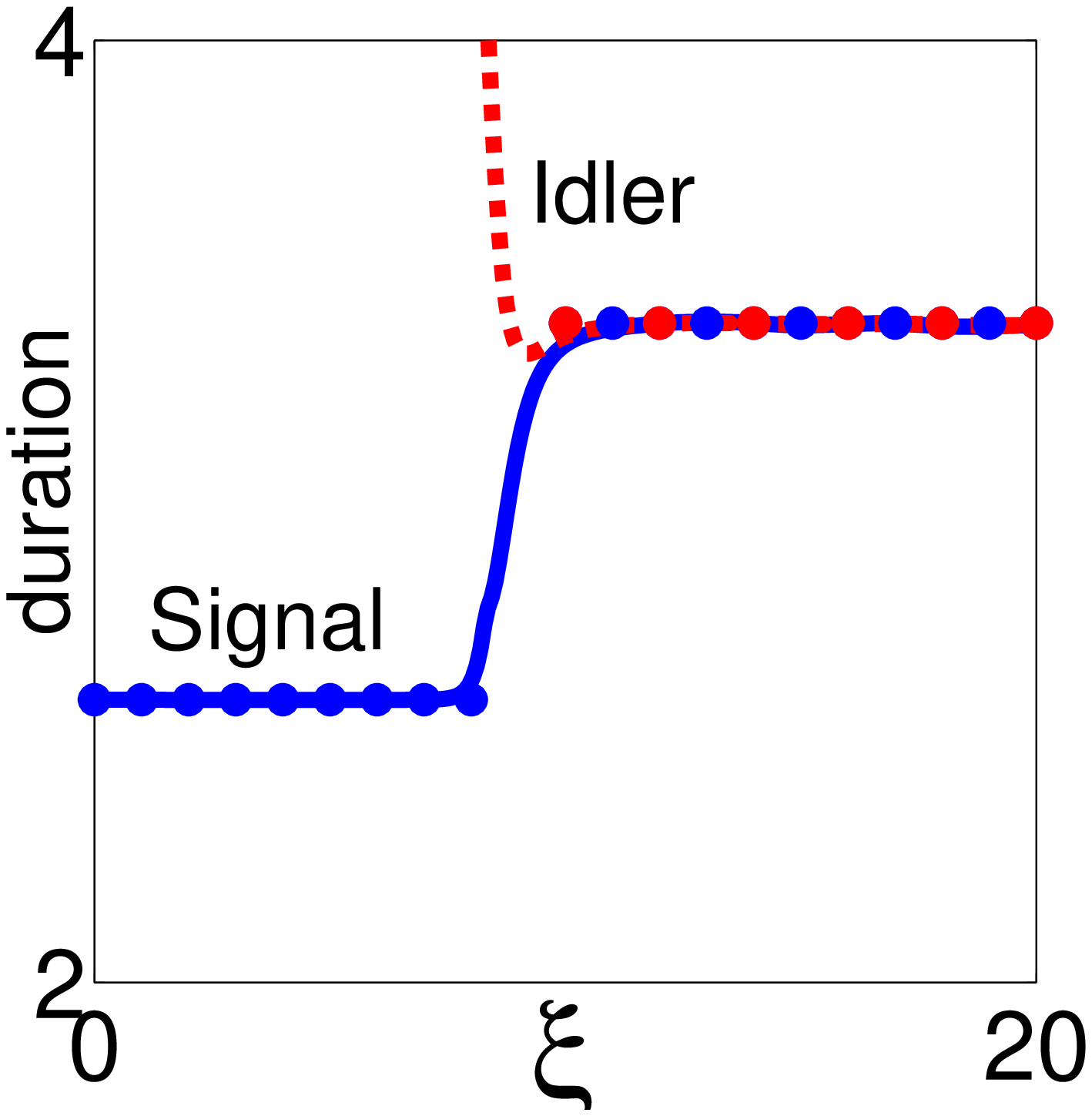}
     \includegraphics[width=3.9cm]{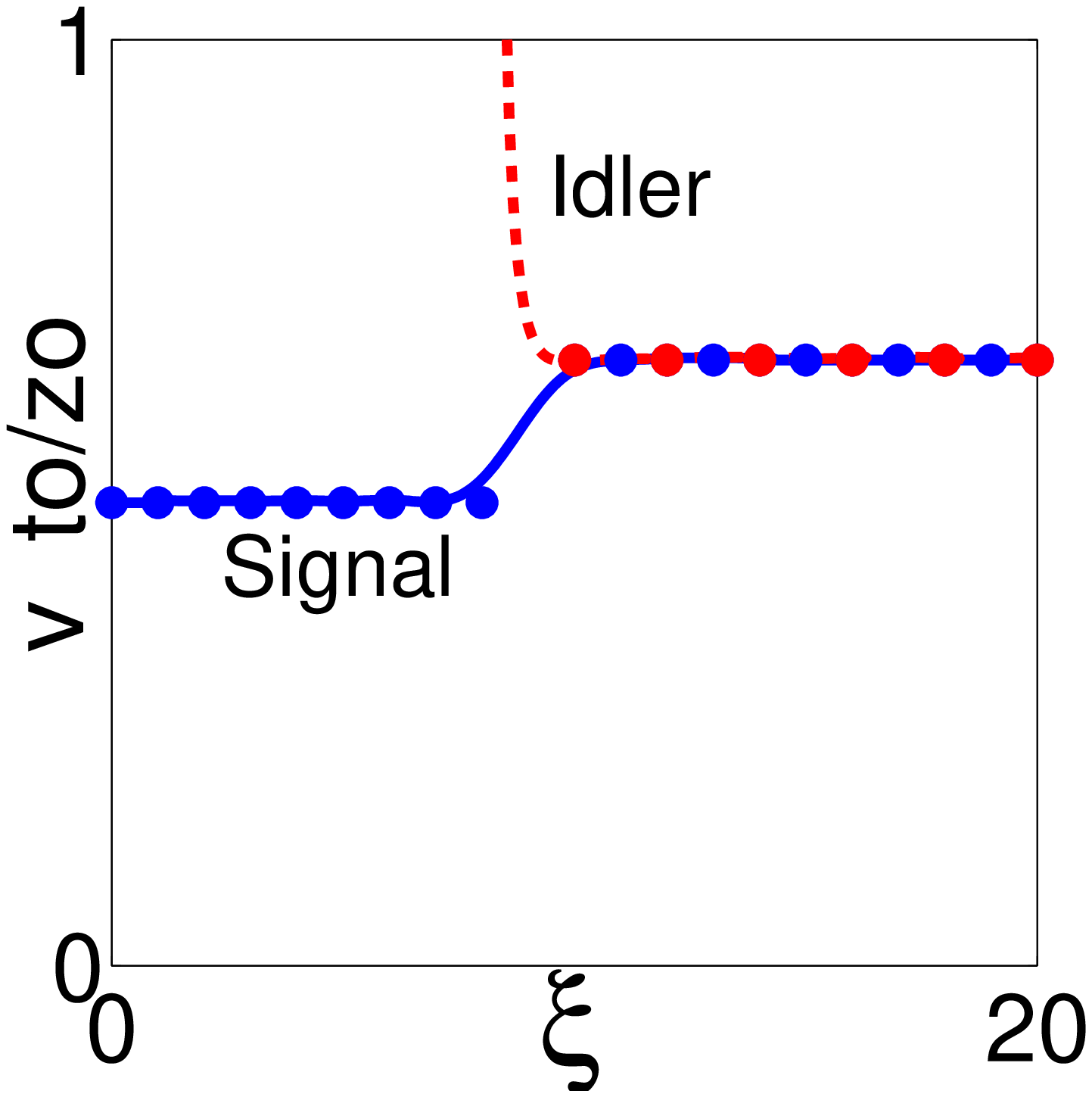}
    \end{center}
     \caption{Numerical evolution (lines) and theoretical predictions
     (circles) of energy, pulse
     duration and velocity of idler and signal waves reported in Fig.\ref{bbk}. }\label{teor}
\end{figure}

As an example, in Fig. \ref{bbk} the input TWRI soliton at
frequency $\omega_2$ is described by Eqs. (\ref{simult-}) with
$p=1.3,k=0,q=0,a=0$, and the background control amplitude is
$C=1.7$. After the collision with the CW background, the above
equations predict that the generated TWRIS is again described by
Eqs. (\ref{simult-}), with
$\overline{p}=1.3,\overline{k}=0,\overline{q}=0,$ and
$\overline{a}=1.2$. The accuracy of this prediction is well
confirmed by its comparison with the numerical solutions of the
TWRI Eqs. (\ref{3wri}). Indeed, Fig. \ref{teor} compares the
numerical with the analytical evolutions (along the crystal length
$\xi$) of the energy, the pulse duration and the velocity of the
idler and signal pulses which correspond to the case shown in Fig.
\ref{bbk}. We performed further extensive numerical simulations,
which confirmed the general validity of the above described
adiabatic transition model for TWRIS generation upon collision
with a CW background.

Indeed, by increasing or decreasing the CW background amplitude
$|C|$ in the range $[0,p \sqrt{\delta_2\delta_3}]$, we observed
that stable TWRISs with different velocity, duration and energy
distributions may be adiabatically shaped. The important
consequence of this result is that, by means of Eqs.
(\ref{simult-})--(\ref{defteor}), we may analytically predict and
control the characteristics of the generated idler pulse (namely,
its velocity, time duration and energy) simply as a function of
the intensity level of the CW pump. Moreover, we would like to
emphasize that the stability of the whole SF idler conversion
process is ensured by the underlying stability of the generated
TWRIS \cite{conforti06}.

\section{Conclusions}
In conclusion, we demonstrated the parametric SF conversion of an
ultra-short pulse from the mixing of an ultra-short optical pulse
with a quasi-continuous wave control in quadratic nonlinear
crystals in the presence of GVM. We analytically showed that the
intensity, time duration and group velocity of the generated
pulses may be controlled in a stable manner by simply adjusting the
intensity level of the background pump.

\end{document}